\documentclass{ws-procs975x65}

\begin{document}

\title{\uppercase{Bounds on Spontaneous Collapse model of Quantum Mechanics from
  formation of CMBR and Standard Cosmology}}

\author{\uppercase{Suratna Das}$^*$, \uppercase{Kinjalk Lochan} }

\address{Tata Institute of Fundamental Research, Mumbai 400005, India\\
$^*$E-mail: suratna@tifr.res.in}

\author{\uppercase{Angelo Bassi}}

\address{Department of Physics,
University of Trieste, Strada Costiera 11, 34151 Trieste, Italy.
\\ IIstituto Nazionale di Fisica Nucleare, Trieste Section, Via Valerio 2, 34127 Trieste,
Italy.}

\begin{abstract}
Continuous Spontaneous Localization (CSL) model of Quantum Mechanics
modifies Schr\"{o}dinger equation by adding non-linear stochastic
terms due to which the total energy of a system increases with a
constant rate which is proportional to the collapse rate
$\lambda$. Thus applying CSL model to cosmological scenarios can
change the thermal behaviour of the particles during evolution and we
will put constraints on $\lambda$ by considering several cosmological
scenarios.
\end{abstract}


\bodymatter

\section{Introduction}
The remarkable difference between the probabilistic dynamics of
microscopic quantum particles and the deterministic nature of
macroscopic classical objects remains a long-standing puzzle since the
advent of Quantum Mechanics. Models of spontaneous collapse of
wavefunctions \cite{Bassi, Bassi:2012bg} is one attempt to unify these
two apparently different dynamics by modifying the Schr\"{o}dinger
equation with additional non-linear terms which breakdown the
superposition of wavefunctions at macroscopic level yielding
deterministic dynamics for classical objects. These non-linear terms
also act as amplification mechanism such that the modified dynamics
only affect the macroscopic dynamics preserving the probabilistic
nature of the microscopic ones. Among many attempts to construct
models of spontaneous collapse of wavefunction Continuous Spontaneous
Localization (CSL) \cite{Ghirardi:1989cn, PS} model is one which will
be of our main focus in this discussion. This model has two free
parameters, the collapse rate $\lambda$ and the width of the
localization $r_c$, whose theoretical origins are yet to be
determined. Also, in its present form the CSL mechanism is only
applicable to non-relativistic particles. The consequence of adding
non-linear terms to Schr\"{o}dinger equation is that the energy of the
system increases with time and the energy density of a system
containing a species $s$ increases with a rate as
\begin{eqnarray}
\left.\frac{\partial \varepsilon_s}{\partial t}\right.=\frac{3\alpha\lambda\hbar^2}{4m_{s}}n_{s}\, ,
\label{heating-rate}
\end{eqnarray}
where $\alpha=\frac{1}{r_c^2}$, $m_s$ is the mass of a
non-relativistic particle of that species and $n_s$ is the number
density. In generic CSL scenario the collapse rate $\lambda$ is taken
to be independent of the mass of the non-relativistic particle. But in
a variant scenario it is considered to be proportional to the square
of the mass of the particle under consideration as
\begin{equation}
 \lambda(m)=\lambda_0\left(\frac{m}{m_N}\right)^2,
\end{equation}
where $m_N$ is the mass of a nucleon. This scenario shows that the
collapse rate can be different for different species and the collapse
rate is higher for heavier ones. We will investigate the cosmological
implications of such a feature of CSL in this talk and put constraint
on $\lambda$ from various cosmological facts and observations.

\section{Effects of CSL heating on different cosmological scenarios and bounds on $\lambda$}

We will now investigate the cosmological consequences of heating up of 
non-relativistic particles in the cosmic soup due to CSL mechanism. 

1. First of all, as the total energy density of the particles change with
a constant rate duo to CSL heating the evolution of energy density of
non-relativistic matter in the radiation and matter dominated eras
changes which can be given as \cite{Lochan:2012di}
\begin{eqnarray}
\rho_{M_f}^{\rm RD}&=&\exp\left\{K_{\rm CSL, RD}\left[\frac{1}{(1+z_f)^2}-\frac{1}{(1+z_i)^2}\right]\right\}\left(\frac{1+z_f}{1+z_i}\right)^3\rho_{M_i},\\
\rho_{M_f}^{\rm MD}&=&\exp\left\{K_{\rm CSL, MD}\left[\frac{1}{(1+z_f)^\frac32}-\frac{1}{(1+z_i)^\frac32}\right]\right\}\left(\frac{1+z_f}{1+z_i}\right)^3\rho_{M_i},
\label{rho-rd}
\end{eqnarray}
where $K_{\rm CSL,
  RD}\approx\sum_{s=e,p}\frac{3\times10^{-9}\lambda\hbar^2\alpha}{8m_{s}m_pc^2\Omega_{R_0}^{\frac12}H_0}$
and $K_{\rm CSL,
  MD}\approx\sum_{s=e,p}\frac{\lambda\hbar^2\alpha}{6m_{s}m_pc^2\Omega_{M_0}^{\frac12}H_0}$. We demand that the CSL mechanism will not effect the standard
evolution of the non-relativistic matter during radiation and matter
dominated phases of the Universe.

2. Next we see that the precise blackbody spectrum of CMBR indicates
that the radiation and matter were in perfect thermal equilibrium
before decoupling at $z\approx1100$. Compton scattering is the most
significant process in thermalization of cosmic plasma once the photon
number non-conserving processes die out at $z\approx10^6$. The time of
thermalization due to Compton scattering is
\begin{eqnarray}
t_{\rm C}=\frac{3m_ec}{4\sigma_T\varepsilon_\gamma}\simeq \, 7.63\times10^{19}(1+z)^{-4}\,{\rm s}.
\end{eqnarray}
We demand that the CSL heating rate should be much smaller than the
Compton thermalization rate so that the radiation and matter in the
cosmic soup can attain thermal equilibrium to yield a perfect
blackbody spectrum.

3. Energy injections at early epochs in the cosmic plasma via direct
heating of particles or by injecting high frequency photons relaxes
the Planck spectrum of the thermalized photon density to a
Bose-Einstein distribution with a non-zero chemical potential
$\mu$. This is known as the $\mu-$type spectral distortion of
CMBR. COBE/FIRAS \cite{cobe} puts an upper bound on such a distortion
of the blackbody spectrum as $\mu\leq 9\times 10^{-5}$ whereas an
upcoming experiment PIXIE \cite{Kogut:2011xw} can constrain it up to
$\mu\sim5\times10^{-8}$. CSL heating can yield a $\mu-$type distortion
in the CMBR of the amount \cite{Lochan:2012di} :
\begin{eqnarray}
\delta\mu=\frac{3}{2.143}\frac{\delta\varepsilon_{\rm CSL}}{\varepsilon_\gamma}.
\end{eqnarray}
One can constrain $\lambda$ from the upper-bounds on $\mu-$type
distortions from these two experiments.

4. If energy releases in the early universe during $z\leq5\times10^4$
makes the temperature of the electrons greater than that of the
photons then low-energy photons will be upscattered in frequency via
Compton scattering causing a deficit of photons in the low frequency
regime and an increment of photons at high frequencies in comparison
with a standard blackbody spectrum. Such a distortion is known as the
$y-$type distortion of the CMBR spectrum. COBE/FIRAS \cite{cobe} puts
an upper bound on such a distortion of the blackbody spectrum as
$y\leq 1.5\times 10^{-5}$ whereas an upcoming experiment PIXIE
\cite{Kogut:2011xw} can constrain it up to $y\leq10^{-8}$. One can
calculate the amount of $y-$type distortion due to CSL heating of
non-relativistic particles by \cite{Lochan:2012di}
\begin{eqnarray}
\delta y=\frac14\frac{\delta\varepsilon}{\varepsilon}.
\label{y-param1}
\end{eqnarray}
Hence, upper-bounds on such a distortion from the two experiments
mentioned above can also constrain the parameter $\lambda$ of CSL.

We consider all these above mentioned scenarios and put constraints on
$\lambda$ parameter of CSL which we summarize in the following table
\begin{table}[h]
\caption{Bounds on CSL strength parameter}
\centering
\small{
\begin{tabular}{c c c}
\hline\hline 
Case &$\lambda$ (in s$^{-1}$) & $\lambda_0$ (in s$^{-1}$)\\ 
\hline 
\small{Bounds from RD era of standard cosmology} & \small{$\ll5\times10^{10}$} &\small{$\ll10^{14}$}\\ 
Bounds from MD era of standard cosmology & $\ll4\times10^{-4}$ & $\ll0.7$ \\ 
Bounds from comparing rates of Compton scattering and CSL heating &2$\times10^3$ &$3\times10^6$\\ 
Bounds from COBE/FIRAS observation of $\mu-$distortion & 70 & $10^{5}$ \\ 
Bounds from PIXIE future observation of $\mu-$distortion &$4\times10^{-2}$ & 74\\ 
Bounds from COBE/FIRAS observation of $y-$distortion &$8\times10^{-5}$ & $0.14$\\ 
Bounds from PIXIE future observation of $y-$distortion & $5\times10^{-8}$ &$10^{-4}$ \\
\hline
\end{tabular}}
\label{lambda-table}
\end{table}



\begin{thebibliography}{99}
\bibitem{Bassi}
 A.~Bassi and G.~C.~Ghirardi,
  {\em Phys.\ Rept.\ } {\bf 379}, 257 (2003)
  [arXiv:quant-ph/0302164].
\bibitem{Bassi:2012bg} 
  A.~Bassi, K.~Lochan, S.~Satin, T.~P.~Singh and H.~Ulbricht,
  arXiv:1204.4325 [quant-ph].
\bibitem{Ghirardi:1989cn} 
  G.~C.~Ghirardi, P.~M.~Pearle and A.~Rimini,
  {\em Phys.\ Rev.\ }  {\bf A42}, 78 (1990).
\bibitem{PS} 
P.~Pearle and E.~Squires, {\em Phys.\ Rev.\ Lett.\ }{\bf 73} (1994).
\bibitem{Lochan:2012di} 
  K.~Lochan, S.~Das and A.~Bassi,
  {\em Phys.\ Rev.\ } {\bf D86}, 065016 (2012)
  [arXiv:1206.4425 [astro-ph.CO]].
\bibitem{cobe}
J.~C.~Mather {\it et al.},
  {\em Astrophys.\ J.\ }  {\bf 420}, 439 (1994).
\\
D.~J.~Fixsen, E.~S.~Cheng, J.~M.~Gales, J.~C.~Mather, R.~A.~Shafer and E.~L.~Wright,
  {\em Astrophys.\ J.\ }  {\bf 473}, 576 (1996)
  [arXiv:astro-ph/9605054].
\bibitem{Kogut:2011xw}
  A.~Kogut {\it et al.},
  {\em JCAP} {\bf 1107}, 025 (2011)
  [arXiv:1105.2044 [astro-ph.CO]].
\end{thebibliography}
\end{document}